\documentclass[aps, prl, twocolumn, ,amsmath, amssymb, verbatim]{revtex4}

\usepackage{graphicx}
\usepackage[dvips]{hyperref}
\usepackage{textcomp}

\begin{document}

\title{Quantum mechanics is based on a relativity principle}
\author{L\'eon Brenig}
\affiliation{Universit\'e Libre de Bruxelles. CP.231. 1050 Brussels.
Belgium.}\label{I1}
\date{16 July 2006}

\begin{abstract}
Non-relativistic quantum mechanics is shown to emerge from classical
mechanics through the requirement of a relativity principle based
on special transformations acting on position and momentum uncertainties.
These transformations are related to dilatations of space variables
provided the quantum potential is added to the classical Hamiltonian
functional. The Schr\"odinger equation appears to have a nonunitary
and nonlinear companion acting in another time variable. Evolution
in this time seems related to the state vector reduction
\end{abstract}
\newcommand\revtex{Rev\TeX}
\maketitle

Unlike special relativity and the Einstein's theory of gravitation,
quantum mechanics is not considered to be based on a principle of
relativity. The formers require the invariance of the mathematical
representation of the laws of Nature under transformations relating
so-called observers or space-time frames of reference. The demand
that classical mechanics becomes invariant under these transformations
entails modifications of the fundamental laws of physics.

Our claim in this Letter is that quantum mechanics is also a relativity theory in
the sense that this theory emerges out of classical mechanics from
the condition of invariance of the laws of mechanics under a new
group of frame transformations. This group acts on the precision
of position and momentum measurements of the observer. Observers
or frames of references are characterized not only by the position
of their space and time origins, direction of their axis and relative
velocities but also by the accuracy or resolution of their instruments.
Such an approach to quantum mechanics has not been contemplated
often in modern physics. The most notorious exception is, to our
knowledge, the interesting work of Nottale \cite{Nottale93},\cite{Nottale
96},\cite{Nottale 04}. The theory reported here, however, differs
fundamentally from that developed by Nottale. 

This difference is mainly due to the fact that we do not attribute a
fractal character to space-time in contrast with this researcher. 
In Nottale's work, this fundamental fractality is inferred from the 
discovery by Abbott and Wise that the trajectory of a quantum free particle
has fractal dimension 2 \cite{Abbott}. Nottale relates this characteristics
to a fundamental fractal property of space-time. Due to this fractality, he is led 
to consider scale transformations. These transformations are, however,
very different from ours. Moreover, the meaning of a fractal trajectory is somewhat
vague in the sense that many aspects of quantum mechanics point to the
vacuity of the concept of trajectory. In contrast, our work does not invoke such
concept and remains in the usual Copenhague interpretation frame. There is,
however, a possible articulation with Nottale's work in the sense that starting with
a Galilean space-time and requiring only invariance of physics under a well-defined 
relativity group acting on measurements precision, we find a supplementary
time dimension. Hence, "trajectories" are defined by two temporal real parameters
and can be viewed as 2-dimensional objects. But this affects the topological dimension
and not the fractal dimension. This is, perhaps, the limit between the two approaches.

Due to the concision inherent to a Letter the demonstrations of many results reported here
are only outlined, though, all of them can be retraced by the reader. A more complete
article will be submitted elsewhere containing also generalization to a particle in an
exterior potential and to systems of N particles interacting via a binary potential. 
 
Let us consider a non-relativistic free particle described by the
Schr\"odinger equation. The transformations on the uncertainties
we are proposing are the following
\begin{gather}
{\Delta x'}^{2}= e ^{-\alpha }{\Delta x}^{2}
\\{\Delta p'}^{2}= e ^{-\alpha }{\Delta p}^{2}+\frac{\hbar ^{2}}{4}\left(
e ^{\alpha }- e ^{-\alpha }\right) \frac{1}{{\Delta x}^{2}}
\end{gather}

where ${\Delta x}^{2}$ is the Fisher dispersion of the position
probability distribution (p.d.) $\rho$(x). It is given by 
\begin{equation}
{\Delta x}^{2}=\frac{1}{2\int {\left| \nabla \rho ^{1/2}( x) \right|
}^{2}d^{3}x}
\end{equation}

where the denominator is the Fisher information associated to the
p.d. $\rho$ \cite{Hall}.

Its relation with the position mean square deviation ${\sigma _{x}}^{2}$ is
given by the Cram\'er-Rao inequality 
\begin{equation}
{\sigma _{x}}^{2}\geq {\Delta x}^{2}
\end{equation}

a classical result in theoretical statistics obtained from the Schwartz
inequality \cite{Cox} .

The quantity ${\Delta p}^{2}$ is the mean square deviation of momentum
calculated with the standard quantum algorithm for the expectation
of any function of the momentum operator.
The group property of the above transformations is easy to establish.
From here on, we consider it as the relativity group relating the various 
observers and under which the laws of physics should be covariant.

Multiplying equation (1) by equation (2) one gets
\begin{equation}
{\Delta x'}^{2}{\Delta p'}^{2}= e ^{-2\alpha }{\Delta x}^{2}{\Delta
p}^{2}+\frac{\hbar ^{2}}{4}\left( 1- e ^{-2\alpha }\right) 
\end{equation}

The parameter $\alpha$ is any real number. The one-parameter continuous
group structure of the set\ \ of these transformations is easy to
prove. Furthermore, when $\alpha$$ \rightarrow $+$ \infty $, ${\Delta
x'}^{2}{\Delta p'}^{2}$$ \rightarrow $$\frac{\hbar ^{2}}{4}$. If
${\Delta x}^{2}{\Delta p}^{2}$ is already equal to $\frac{\hbar
^{2}}{4}$ then the product ${\Delta x'}^{2}{\Delta p'}^{2}$ keeps
the value $\frac{\hbar ^{2}}{4}$ for any value of $\alpha$. For
$\alpha$$ \rightarrow $-$ \infty $, ${\Delta x'}^{2}{\Delta p'}^{2}$$
\rightarrow $+$ \infty $ for any value of ${\Delta x}^{2}{\Delta
p}^{2}$$ \geq $$\frac{\hbar ^{2}}{4}.$ Of course, since the Cram\'er-Rao
inequality (4) guarantees that\ \ ${\sigma _{x}}^{2}$${\Delta p}^{2}$$
\geq $${\Delta x}^{2}{\Delta p}^{2}$, all these asymptotic results
are lower boundaries for the values of ${\sigma _{x}}^{2}$${\Delta
p}^{2}$ and its transformations.

These remarkable properties bear some similarities with the Lorentz
transformations. In the same way the velocity of light constitutes
an upper value for the velocities of material bodies, the constant $\frac{\hbar ^{2}}{4}$
represents a lower limit value for the product
of uncertainties ${\Delta x}^{2}{\Delta p}^{2}.$

The choice of the above transformations (1), (2) as relativity group
for the laws of\ \ physics imposes a radical modification of the
laws of dynamics that corresponds to the passage from classical
to quantum mechanics as we prove now. 

Let us consider the classical mechanical description of a free non-relativistic
particle of mass m. We describe it in a field canonical framework
\cite{Sudarshan83} by introducing at the initial time the p.d. $\rho$(x)
of an ensemble of identical non-interacting particles. This function
together with the classical action of the particle, s(x), are the
basic field variables of the formalism. The time evolution of any
functional of type
\begin{equation}
\mathcal{A} = \int d^{3}x F( x, \rho , \nabla \rho , \nabla \nabla
\varrho , ..., s, \nabla s, \nabla \nabla s, ...) 
\end{equation}

of the two variables $ \rho $ and s\ \ that is at least once functionally
differentiable in terms of $ \rho $ and s is given by
\begin{equation}
\partial _{t}\mathcal{A} = \left\{ \mathcal{A},\mathcal{H}_{\mathrm{cl}}\right\}
\end{equation}

where
\begin{equation}
\mathcal{H}_{\mathrm{cl}}= \int d^{3}x\ \ \frac{\rho \left| \nabla
s|^{2}\right. }{2m}
\end{equation}

is the classical Hamiltonian functional and
\begin{equation}
\left\{ \mathcal{A},\mathcal{B}\right\}  = \int d^{3}x\ \ [\frac{\delta
\mathcal{A}}{\delta \rho ( x) }\frac{\delta \mathcal{B}}{\delta
s( x) }- \frac{\delta \mathcal{B}}{\delta \rho ( x) }\frac{\delta
\mathcal{A}}{\delta s( x) }]
\end{equation}

where $\frac{\delta }{\delta \rho ( x) }$ and $\frac{\delta }{\delta
s( x) }$ are functional derivatives. The above functional Poisson
bracket endows the set of functionals of type (6) with an infinite
Lie algebra structure $ \mathbb{G}$.

Any functional of $ \mathbb{G}$, and $\mathcal{H}_{\mathrm{cl}}$
is one of them, generates a one-parameter continuous group of transformations
. The time transformations are generated by $\mathcal{H}_{\mathrm{cl}}$.
Equation (7) when applied to $\rho$(x) and s(x) respectively, yields
the continuity equation and the Hamilton-Jacobi equation
\begin{gather}
\partial _{t}\rho  = -\nabla .\left( \rho \frac{\nabla s}{m}\right)
\\\partial _{t}s =- \frac{{\left| \nabla s\right| }^{2}}{2m}
\end{gather}

where the gradient $\nabla s$ is the momentum of the particle.

Now let us consider the group of space dilatations and its action
on $\rho $ and s
\begin{equation}
\rho '\left( x\right) = e ^{\frac{3\alpha }{2}}\rho (  e ^{\frac{\alpha
}{2}}x) , s'\left( x\right) = e ^{-\alpha }s(  e ^{\frac{\alpha
}{2}}x) 
\end{equation}

where $\alpha$ is any real number. Note that these transformations
preserve the normalization of the p.d. $\rho $(x) \cite{Hall}. Clearly,
they also keep the dynamical equations (10) and (11) invariant.

Let us assume that the average momentum of the particle is vanishing.
This corresponds to a particular choice of the frame of reference
but, by no means, reduces the generality of our results. In this
frame, the quadratic mean deviation of the momentum is given by
\begin{equation}
{{\Delta p}_{\mathrm{cl}}}^{2}=\int d^{3}x\ \ \rho \left| \nabla
s|^{2}=2m \mathcal{H}_{\mathrm{cl}}\right. 
\end{equation}

and under transformations (12) becomes
\begin{equation}
{{\Delta p}_{\mathrm{cl}}'}^{2}= e ^{-\alpha }{{\Delta p}_{\mathrm{cl}}}^{2}
\end{equation}

Also, the Fisher dispersion of $ \rho $(x), ${\Delta x}^{2}$, defined
in equation (3) transforms as
\begin{equation}
{\Delta x'}^{2}= e ^{-\alpha }{\Delta x}^{2}
\end{equation}

Not surprisingly, it appears from equation (14) that the classical
momentum uncertainty does not transform like prescribed by equation
(2) above. In view of the physical dimensions of ${{\Delta p}_{\mathrm{cl}}}^{2}$
as defined in equation (13), transformation (14) is expected under
dilatations affecting the position coordinates. It corresponds to
the first term in the right hand side of equation (2). 

Now, let us modify definition (13) of\ \ ${{\Delta p}_{\mathrm{cl}}}^{2}$
by adding a new term proportional to the Fisher information, $\frac{\hbar
^{2}}{2}F $.\ \ This addition along with the use of equation (3)
yields the following quantity
\begin{equation}
{{\Delta p}_{q}}^{2}= \int d^{3}x\ \ \rho ( x) | \nabla s(
x) |^{2} + \hbar ^{2}\int d^{3}x{\left| \nabla {\rho ( x) }^{1/2}\right|
}^{2} 
\end{equation}

We new prove that the above supplementary term restores the relativity
transformation law (2). Let us apply the space dilatation (12) to
this functional. This leads to
\begin{equation}
{{\Delta p'}_{q}}^{2}= e ^{-\alpha }\int d^{3}x\ \ \rho ( x) |
\nabla s( x) |^{2}+ e ^{\alpha }\hbar ^{2}\int d^{3}x {\left| \nabla
{\rho ( x) }^{1/2}\right| }^{2} 
\end{equation}

 Adding and subtracting an appropriate term,\ \ $ e ^{-\alpha }\hbar
^{2}\int d^{3}x {|\nabla {\rho ( x) }^{1/2}|}^{2}$, to the right
hand side of equation (17) allows expressing this equation in terms
of ${{\Delta p}_{q}}^{2}$ as follows
\begin{equation}
{{\Delta p'}_{q}}^{2}= e ^{-\alpha }{{\Delta p}_{q}}^{2}+\frac{\hbar
^{2}}{4}\left(  e ^{\alpha }- e ^{-\alpha }\right) \frac{1}{{\Delta
x}^{2}}
\end{equation}

where we have used again equation (3). This equation is identical
to the transformation law (2). We may, thus, identify ${\Delta p}_{q}$ with
$\Delta p$ , i.e. the quantal momentum uncertainty. Since the Hamiltonian
functional for a free particle is its average kinetic energy, we
have in this frame
\begin{equation}
\mathcal{H}_{q}=\frac{{{\Delta p}_{q}}^{2}}{2m}
\end{equation}

or
\begin{equation}
\mathcal{H}_{q}=\int d^{3}x\ \ \frac{\rho ( x) \left| \nabla
s( x) |^{2}\right. }{2m} +\frac{ \hbar ^{2}}{2m}\int d^{3}x{\left| \nabla {\rho
( x) }^{1/2}\right| }^{2} 
\end{equation}

This is precisely the expected expression of the quantum average
of the energy for a free particle. Clearly, the apparition of the
Planck constant in this derivation is quite artificial. The constant
multiplying the Fisher information in the added term in (16) is
just arbitrary. The Planck value of that constant has been postulated
in order to retrieve quantum mechanics.

The functional $\mathcal{H}_{q}$ generates the quantum time evolution
of any functional $\mathcal{A} $ of the algebra $ \mathbb{G}$ via
equation (7) where $\mathcal{H}_{\mathrm{cl}}$ is to be replaced
by $\mathcal{H}_{q}$. In particular it gives the Schr\"odinger equation
when $ \mathcal{A}$ is just the wave function $\rho ^{1/2} e ^{\mathrm{is}/\hbar
}$. We leave this demonstration to the reader. One of the intermediate
results is the apparition of the quantum potential \cite{Bohm} in
the Hamilton-Jacobi equation (11)
\begin{equation}
\partial _{t}s =- \frac{{\left| \nabla s\right| }^{2}}{2m}+ \frac{
\hbar ^{2}}{2m}\frac{\nabla ^{2}\rho ^{1/2}}{\rho ^{1/2}}
\end{equation}

while the continuity equation for $\rho$ (10) is preserved.

Let us summarize. We have derived the quantum evolution for a free
particle from the requirement that the quadratic uncertainties on
position and momentum should satisfy the relativity transformations laws
(1) and (2). This result is easily generalized to a particle in
an exterior potential or to N particles interacting via a binary
potential. The form in which we obtain quantum mechanics is that
of canonical field theory which has been introduced and studied
from different points of view by various authors \cite{Strocchi},
\cite{Heslot}, \cite{Guerra83}, \cite{Ashtekar}, \cite{Hall2}. None
of these authors, however, derives quantum mechanics from a relativity
principle as we do here. They assume the existence of the quantum
formalism based on Hilbert space and the algebra of observable operators
acting on it, and show that this framework can be derived from a
more general symplectic or canonical field theory and/or from a
variational principle.

One more important question has now to be investigated: That of
the non-invariance of the Schr\"odinger equation under the transformation
(1), (2). It is a well-known fact that this equation is not invariant
under the conformal group and our transformations are, indeed, dilatations
of position coordinates.\ \ Let us first remark that the generator
of transformations (1) and (2) in the algebra $ \mathbb{G}$ is the
functional
\begin{equation}
\mathcal{S}=\int d^{3}x\ \ \rho ( x) s( x) 
\end{equation}

This can readily be verified by exponentiating the infinitesimal
transformation
\begin{equation}
{{\Delta p'}_{q}}^{2}={{\Delta p}_{q}}^{2}+\delta \alpha \left\{
{{\Delta p}_{q}}^{2},\mathcal{S}\right\} 
\end{equation}

where the bracket is still the one defined in equation (9). So doing,
one gets the same expression as equation (17) or (2). Let us now
define the following new functional
\begin{equation}
\mathcal{K}_{q}\equiv \left\{ \mathcal{S},\mathcal{H}_{q} \right\}
=\int d^{3}x\ \ \frac{\rho \left| \nabla s|^{2}\right. }{2m}
- \frac{ \hbar ^{2}}{2m}\int d^{3}x\ \left| \nabla \rho ^{1/2}\right| ^{2}
\end{equation}

and let us apply the group generated by $ \mathcal{S}$ on both $\mathcal{H}_{q}$ and
$\mathcal{K}_{q}$. An easy calculation yields
\begin{equation}
{\mathcal{H}'}_{q}=\mathrm{cosh\alpha }\ \mathcal{H}_{q}-\mathrm{sinh\alpha
} \ \mathcal{K}_{q}, {\mathcal{K}'}_{q}=-\mathrm{sinh\alpha } \ \mathcal{H}_{q}+\mathrm{cosh\alpha
} \ \mathcal{K}_{q}
\end{equation}

This is due to the fact that $\{\mathcal{S},\mathcal{K}_{q} \}$
is equal to $\mathcal{H}_{q}$. These transformations are isomorphic
to 2-D Lorentz\ \ transformations.

Since $\mathcal{K}_{q}$ only differs from $\mathcal{H}_{q}$ by the
sign of the quantum potential, the group it generates is parametrized
by a new time parameter, $\tau$. Any functional $ \mathcal{A}$ of
$ \mathbb{G}$ can be considered as a function of both t and $\tau$
and its evolution in both times is given by
\begin{equation}
\partial _{t}\mathcal{A} = \left\{ \mathcal{A},\mathcal{H}_{q}\right\}
, \partial _{\tau }\mathcal{A} = \left\{ \mathcal{A},\mathcal{K}_{q}\right\}
\end{equation}

Note also that both generators tend to $\mathcal{H}_{\mathrm{cl}}$ for
$\hbar $$ \rightarrow $0, i.e. both times variables become identical
in the\ \ classical limit. For finite value of $\hbar $ the transformations
(25) induce Lorentz-like transformations in the plane (t, $\tau$)
\begin{equation}
t'=\mathrm{cosh\alpha }\ \ t+\mathrm{sinh\alpha }\  \tau , \tau '=\mathrm{sinh\alpha
}\ \ t+\mathrm{cosh\alpha } \ \tau 
\end{equation}

Also, the remarkable property\ \ that $\mathcal{H}_{q}+i$$\mathcal{K}_{q}$
is a holomorphic function of t+i$\tau$ is easily shown.

The transformations generated by $ \mathcal{S}$ mix the two time
evolutions, hence, only system (26) is covariant but not the individual
equations constituting it. Let us now consider the case where $\mathcal{A}$
is the wave function $\psi$\ \ given by $\rho ^{1/2} e ^{\mathrm{is}/\hbar
}$. As stated above its evolution equation in variable t is linear
and is Schr\"odinger's equation. In time $\tau$, however, the equation
is nonlinear
\begin{equation}
\mathrm{i\hbar }\partial _{\tau }\psi =-\frac{\hbar ^{2}}{2m}\nabla
^{2}\psi +\frac{\hbar ^{2}}{m}\psi \frac{\nabla ^{2}\left| \psi
\right| }{\left| \psi \right| }
\end{equation}

The nonlinear Schr\"odinger equation obtained here is not a newcomer
in physics. It has been envisaged, though in the time t variable
and in different contexts, by several authors \cite{Guerra 1}, \cite{Vigier},
\cite{Smolin}. It belongs to the class of Weinberg's nonlinear Schr\"odinger
equations \cite{Weinberg}. This equation admits a nonlinear superposition
principle \cite{Auberson}. It has been studied, always in the usual
time variable, as a member of the general class of nonlinear Schr\"odinger
equations obtained under the so-called nonlinear gauge transformations
introduced by Doebner and Goldin \cite{Doebner}. The evolution generated
by this equation in\ \ our new time dimension is nonunitary as $
\mathcal{K}_{q}$ can not be reduced to the quantum average of a Hermitian
operator. One easily shows also that together with the functionals
generating translations, rotations and Galilean boosts, $\mathcal{K}_{q}$ constitutes
a field canonical representation of the Galilei algebra. A potentially
important property is that equation (29) implies the continuity
equation for the p.d. $\rho$.

Moreover, the functional $ \mathcal{S}$ is a Lyapunov-like function
for this equation as
\begin{equation}
\partial _{\tau }\mathcal{S}=\left\{ \mathcal{S},\mathcal{K}_{q}\right\}
=\mathcal{H}_{q}\geq 0
\end{equation}

This property is related to the fact that while $\partial _{t}$${\Delta
x}^{2}$$ \geq $0 and $\partial _{t}$${\Delta p}^{2}$=0, the product\ \ ($\partial
_{\tau }$${\Delta x}^{2})$($\partial _{\tau }{\Delta p}^{2}$) is
always negative. This is reminiscent of the process of state vector
reduction in position measurement in which ${\Delta x}^{2}\rightarrow
0$ while ${\Delta p}^{2}\rightarrow +\infty $, or conversely if
one is measuring momentum. Would this evolution correspond to the
nonunitary process that authors like R.Penrose \cite{Penrose} are
invoking for the description of the collapse of the wave function?
The difference with these approaches lies, at least, in the fact
that they always consider the reduction process in the usual time.

Several directions of generalization of our theory can be envisaged. One would
be abandoning global invariance with respect to transformations (1) 
and (2) and requiring only local invariance.
This could lead to the discovery of a new gauge field. Another orientation
would be the extension of the above approach to the case of Klein-Gordon
and Dirac equations, and more generally to quantum field theory with, perhaps,
important consequences at the level of the general unified theory including gravitation.
 
The author wants to dedicate this Letter to his master and friend
Dr.R.Balescu who suddenly deceased during this work and who encouraged
him in this approach. He wants also to thank Drs. C.George, J.Reignier,
Y.Elskens, I.Veretennicoff, G.Barnich, R.Lambiotte, C.Schomblondt
and Mr.F.Ngo for their many relevant comments on this work.

Email address: lbrenig@ulb.ac.be

\end{document}